\documentclass[namedreferences,hyperref,optionalrh,solaromanenum]{spr-sola}
\usepackage{lmodern}

\usepackage{float}
\usepackage{amsmath, amssymb}
\usepackage{graphicx}                    % For eps figures, newer & more powerfull
\usepackage{color}                       % For color text: \color command
\usepackage{hyperref}

\newcommand{\aap}{{\it Astron. Astrophys.}}

\newcommand{\apjl}{{\it Astrophys. J. Lett.}}
\newcommand{\apjs}{{\it Astrophys. J. Suppl. Ser.}}

\chardef\us=`\_

\begin{document}

\begin{frontmatter}

\title{Origin of Coronal Extreme Ultraviolet Shockwaves without a Coronal Mass Ejection Event}

%%%%%%%%%%%%%%%%%%%%%%%%%%%%%%%%%%%%%%%%%%%%%%%%%%%
%% Authors Names
%
\author[addressref={address_id1,address_id2},corref,email={Robert.Bush@tufts.edu}]
       {\inits{R.N. }\fnm{Robert }\snm{Bush}\orcid{0009-0006-2737-8574}}
       
\author[addressref={address_id2},email={john.stefan@njit.edu}]
       {\inits{J.T. }\fnm{John }\snm{Stefan}\orcid{0000-0002-5519-8291}}

\author[addressref={address_id2,address_id3},email={alexander.g.kosovichev@njit.edu}]
       {\inits{A.G. }\fnm{Alexander }\snm{Kosovichev}\orcid{0000-0003-0364-4883}}

%%%%%%%%%%%%%%%%%%%%%%%%%%%%%%%%%%%%%%%%%%%%%%%%%%%
%% Runningheads
%
%\runningauthor{}
%\runningtitle{}

%%%%%%%%%%%%%%%%%%%%%%%%%%%%%%%%%%%%%%%%%%%%%%%%%%%
%% Affilations 
%% id shold be the same with \author addressref value.
\address[id={address_id1}]{Department of Physics and Astronomy, Tufts University, Medford, MA 02155}
\address[id={address_id2}]{Center for Computational Heliophysics, Department of Physics, New Jersey Institute of Technology, Newark, NJ 07102}
\address[id={address_id2}]{NASA Ames Research Center, Moffett Field, Mountain View, CA 94040}

%%%%%%%%%%%%%%%%%%%%%%%%%%%%%%%%%%%%%%%%%%%%%%%%%%%
%%% Abstract 
\begin{abstract}\label{sec:abstract}

A leading theory of sunquake generation involves flare-accelerated particles depositing energy into the photosphere. Simulations of sunquake excitation suggest co-excitation with wavefronts propagating in the corona, similar to large-scale coronal propagating fronts (LCPFs), and also generate Moreton-Ramsey waves in the chromosphere. To investigate observational evidence for the particle-driven mechanism in LCPFs, we compare populations of events associated with and without coronal mass ejections (CMEs). CMEs are known to generate coronal shock waves also observed in EUV emission.  We employ visual inspection of flare events that generate LCPFs using Atmospheric Imaging Assembly (AIA) and Large Angle and Spectrometric Coronagraph (LASCO) coronagraph images to find that the large-scale coronal waves associated with CMEs propagate noticeably faster. Then we examine “standalone flare events” (those that generate coronal waves without CMEs), using soft X-ray (SXR) data from the GOES satellite and focusing on characteristics related to magnetic energy release rate. This reveals that such standalone or confined flares differ from sunquake flares: they are less impulsive and energetic than sunquake flares. However, they are more impulsive but less energetic than LCPF-associated flares with a CME. In particular, coronal waves accompanied by CMEs exhibit significantly higher volume emission measures, suggesting a different generation mechanism.
\end{abstract}

%%%%%%%%%%%%%%%%%%%%%%%%%%%%%%%%%%%%%%%%%%%%%%%%%%%
%% Keywords
%

\end{frontmatter}
%-------------------------------------------------

%%%%%%%%%%%%%%%%%%%%%%%%%%%%%%%%%%%%%%%%%%%%%%%%%%%
%% Sections
%
\section{Introduction}\label{sec:intro}

\cite{Moreton1960} observed propagating waves in the solar chromosphere using H\(\alpha\) images. Later, \cite{Uchida1968} modeled Moreton-Ramsey waves as flare-initiated fast-mode MHD waves propagating in the corona, establishing that coronal shock waves served as the Moreton-Ramsey wave source. These waves appear when the coronal shock interacts with the transition region and generates a chromospheric signature \citep{Patsourakos2009, Stefan2025}, typically observed as variations in the H$\alpha$ line. Moreover, these propagating fronts may also be linked to type II radio bursts, which are radio-wave discharges commonly created when coronal mass ejections (CMEs) accelerate shocks \citep{Nitta2014}. The launch of the Solar and Heliospheric Observatory (SOHO)  \citep[SOHO;][]{Domingo1995} led to the observation of wave-like phenomena observed by the Extreme Ultraviolet Telescope (EIT), denoted as the `EIT waves' \citep{Moses1997, Thompson1998}. They were initially identified as propagating intensity enhancements of coronal extreme-ultraviolet (EUV) emission lines, specifically the Fe XII line around 195\,\AA\, which forms at (approximately greater than) 1.5 MK. Such waves were classified as separate from the coronal shock described by \citet{Uchida1968} due to their strong correlation with CMEs rather than flares \citep{Biesecker2002, Cliver2005, Chen2006}. However, the propagation of EIT waves is consistent with the kinematics of fast-mode MHD waves \citep{Warmuth2001, Warmuth2004, Vrsnak2006} and it has since been found that the differences between the observed EIT waves and coronal disturbances observed by the  Atmospheric Imaging Assembly onboard the Solar Dynamics Observatory (SDO/AIA; \cite{Lemen2012} can be explained by the low cadence and resolution of EIT \citep{Byrne2013}.

\cite{Nitta2013} defined the term "Large-scale Coronal Propagating Fronts," or LCPFs, as a general term to refer to the EIT-wave-like propagating disturbances as observed by AIA. LCPFs were cataloged using visual inspection, exhibiting an angular expanse of (approximately greater than) 45 degrees and propagating at least 200 Mm from the center of the eruption. Their study of a subset containing 86 LCPF events, for which the CME speeds could more reliably be estimated, concluded that the speeds of these LCPFs and associated CMEs were not strongly correlated. The authors suggest that the lack of correlation could be explained by the dependence of the propagation speed on atmospheric conditions.

However, interest was renewed after 3D acoustic simulations using radially-dependent atmospheric heating rates derived from the flare radiative hydrodynamic (RADYN) simulations \citep{Allred2015} were used to realistically excite sunquakes \citep{Sadykov2024}. Extension of these simulations to include the lower corona found evidence of LCPF generation by flare-accelerated particle beams \citep[under review]{Stefan2025}. These disturbances were excited in the atmosphere at the height of maximum energy deposition, typically the upper transition region. The expanding coronal wavefront refracted back across the transition region and propagated into the chromosphere where Moreton-wave-like signatures were generated. The generation of coronal waves by proton or electron beams could help explain events where the Moreton waves, and their respective coronal waves, are observed without a CME; in particular, \citet{Stefan2025} find that  electron beams produce high-amplitude coronal and Moreton waves with hard-to-detect sunquakes. Sunquakes are internal helioseismic waves generated by strong photospheric perturbations during solar flares and were discovered by \cite{Kosovichev1998} using Dopplergrams obtained by the Michelson Doppler Imager (MDI) on SOHO \citep{Scherrer1995}. These acoustic waves travel through the solar convective zone and are observed after they are refracted upwards, spreading out in the photosphere---often anisotropically---following the impulsive phase of a flare. Recent modeling of sunquakes has suggested that particle beams accelerated by flare energies may be the excitation mechanism responsible for helioseismic events \citep{Sadykov2024,Granovsky2025}. In a prior study, at least half of the considered events were consistent with the electron-beam hypothesis \citep{Stefan2020}. 

Characteristics of X-ray emission during solar flares provide important insight into energetics and may reveal the physical processes or mechanisms associated with helioseismic or atmospheric responses. In turn, this may better inform us about the origin of mechanisms driving coronal shockwaves not in the presence of a CME, as these mechanisms may be similar to those of sunquakes. A significant proportion of flare energy contributes to particle acceleration, as a series of recent studies found that the total magnetic energy released during a solar flare/CME event is partitioned as $51\%\pm 17\%$ for electron acceleration, $17\%\pm 17\% $ for ion acceleration, $7\% \pm 14\%$ for CME kinetic energies, and $7\%\pm 17\%$ for direct heating \citep{Aschwanden2019}. Most of these processes can be observed in soft X-ray (SXR) and extreme ultraviolet emissions. CME-driven and particle acceleration generation of coronal waves are inherently different; the former is associated with expanding CME shocks, whereas the latter directly deposits energy into the atmosphere that generates an expanding pressure wave. Thus, we would initially expect the X-ray characteristics of particle-beam-driven flares with coronal waves to resemble flares responsible for helioseismic activity, or at the very least, differ significantly from CME eruptions initiated by flares.

We divide this paper into four main sections. Section \ref{sec:intro} is the introduction and motivation for this research on coronal waves, associated with solar flares. Section \ref{sec:methods} outlines the methodology for collecting and analyzing the data, using observational and computational methods. In Section \ref{sec:results}, we examine the influence of CME presence on the kinematics of the coronal waves and the X-ray characteristics of the associated flares. Section \ref{sec:conclusion} summarizes our results and draws conclusions based on the data and their implications for future studies. 

\section{Methodology}\label{sec:methods}

\subsection{Catalog sourcing}
For analysis, we used several catalogs of sunquakes, coronal waves, and CMEs. The sunquakes cataloged are from 2011--2017 in Solar Cycle 24, as outlined by \cite{Sharykin2020}. There are 114 total, identified using at least one of the following three methods: (1) visually inspected in movies showing time sequences of running-differenced Dopplergrams from the Helioseismic and Magnetic Imager \citep[HMI;][]{Scherrer2012}, filtered in a bandwidth of 5-7 mHz, (2) characteristic ridge patterns found in time-distance (TD) diagrams created from photospheric impacts detected in HMI Dopplergrams, or (3) using the helioseismic holography method \citep{Lindsey2000}, using a Green's function of helioseismic waves to calculate egression acoustic power that corresponds to observed Doppler velocities. 94 were found using at least one of the above methods, and an additional 20 were considered potential candidates identified by the holography method (but not seen in the other two methods). 

The coronal wave catalog was sourced from \cite{Nitta2013}, which contains LCPFs observed by the AIA instrument from 2010 to 2022, with \(\sim  700\) events. We specifically examined the standard SolarSoft \citep{Freeland&Handy1998} movies in AIA's EUV 171\,\AA\, and 193\,\AA\, channels, produced using the running-difference method. Visually, the fronts appear in AIA as approximately radially expanding ripples originating from the flare site. LCPFs are defined as events having an angular expanse of $\gtrsim 45^\circ$ and that propagate at least 200\,Mm away from the center of the associated flare. It should be noted that the speeds of these wavefronts are calculated only for the period from 2010 to 2013 (171 events). This speed estimation was performed by \citet{Nitta2013} and accomplished by fitting a first-order polynomial to the front edge of the most prominent ridge in the distance-time plot. The ridge fits from each of the 24 equally spaced \(15^\circ\)-wide sectors are averaged, assuming uniform uncertainty of 5 Mm, to locate the front using MPFIT \citep{Markwardt2009}.

The CME catalog, containing all CMEs manually identified since 1996, was sourced from the Large Angle and Spectrometric Coronagraph \citep[LASCO;][]{Brueckner1995} onboard SOHO. This CME catalog is generated and maintained at the CDAW Data Center by NASA and The Catholic University of America in cooperation with the Naval Research Laboratory. We mainly examined two aspects of associated CMEs: their linear speed and mass. The linear speed is found by fitting a straight line to the height-time measurements and represents a mean speed across the LASCO field of view. The mass estimates, while very uncertain, have a representative value that remains quasi-constant after traveling the first several solar radii; if the CME fades within this distance, the mass measured at the last time is used. Computationally, these mass values are found using equations from observed Thomson scattering, which measures white light scattered from electrons at the outer edge of the CME plasma \citep{Gopalswamy2024}.

\subsection{Observational methods}
We first visually inspected each of the 171 coronal wave events from April 2010 -- January 2013 to detect whether or not each one had a CME. Before visual inspection, we cross-referenced the coronal wave catalog with the LASCO CME database, considering any coronal wave event to have a corresponding CME if there was a recorded CME less than 90 minutes from the recorded start time of the coronal wave event. This was to provide time for the CME to be properly observed on the coronagraph, since its effects may not be immediately seen after the flare and/or CME. The population of coronal waves from \cite{Nitta2013} propagate out from an origin, initially looking akin to "dropping a pebble in a pond." As they expand, the fronts become increasingly asymmetric (i.e., they do not solely expand radially), where the fronts in some sectors appear darker or brighter, as well as faster or slower, as they cross the solar disk. Then, a visual inspection of both coronagraph and 171\AA/193\AA\, AIA movies during each recorded wave event was performed. The following event was considered to have an accompanying CME if there was a noticeable plasma ejection traveling beyond the solar limb in the same region of both the coronagraph and 171\AA/193\AA\, movies. 

Additionally, we recorded the wave speeds as seen in the LCPF catalog, as well as the recorded closest CME time, mass, kinetic energy, speed, and other remarks (such as if an event was poor, only in C2, or a partial halo was observed) as seen in the LASCO catalog. Other notes were recorded regarding the visual strength and size of an observed CME, especially if it was rather weak, faint, and difficult to observe, or if it was quite noticeable. The motivation for this visual inspection stems from the fact that, as stated by the SOHO LASCO CME catalog team, "in the absence of a perfect automatic CME detector program, the manual identification is still the best way to identify CMEs." This was mainly in an effort to confirm our initial tagging of CME-related events, in which coronal wave/LCPF events were only considered to have a corresponding CME event if the start time of the wave (which corresponds to the closest flare start time) was within 90 minutes of the recorded first C2 appearance of a CME in LASCO. When necessary, we re-tagged events from the programmatically-assigned tag to either having a CME (yes), not having a CME (no), or being inconclusive (maybe) based on this visual inspection. We refer to events without a CME as "isolated".

\subsection{Computational methods}

\subsubsection{Raw GOES SXR flux data}
Next, we analyzed the statistics of flare SXR characteristics of all flares with a GOES class greater than C1.0, which occurred between 2010 and 2022---this is the range of available AIA movies of coronal waves/LCPFs maintained through 2024 by the authors of the LCPF study \citep{Nitta2013}\footnote{The LCPF catalog is publicly available at \href{ https://aia.lmsal.com/AIA_Waves/index.html}{http://aia.lmsal.com/AIA\_Waves/index.html}}. The flare events were tagged in a similar fashion to the coronal wave events; the specific categories are: sunquake-associated flares, non-sunquake flares that produce an isolated coronal wave, non-sunquake flares that produce a CME-related coronal wave, and non-sunquake flares that do not produce a coronal wave.

We paid the most attention to flare X-ray characteristics related to the energy release process as it was previously found that flares that are more impulsive are also helioseismically active \citep{Sharykin2020}. The first characteristic we consider is the impulsive phase duration of each flare, defined as the time interval when the SXR flux was higher than $\max(f_{1-8\,\text{\AA}}/10)$, where $f_{1-8\,\text{\AA}}$ is the GOES SXR flux in the 1--8\,\AA \, channel. The second is the characteristic energy release time estimated from the maximum value of the time-derivative of the natural log of the SXR flux 1--8\,\AA \,channel (i.e. the maximum value of \(f_{1-8\text{\,\AA}}/(df_{1-8\,\text{\AA}}/dt)\)). This quantity indicates how quickly or slowly the SXR emission reaches its peak value. We also examined the maximum value of the SXR-flux and its maximum time derivative during the flare event. Lastly, we examined the total radiated flare energy calculated from the GOES 1--8\,\AA \,flux integrated over the start and end time of the flare under the assumption of isotropic emission.

\subsubsection{Conversion of flux ratios to derive flare temperatures and emission measures}\label{sec:tempEM}

Furthermore, we also considered the important characteristics of flare temperature and emission measure, calculated using the Python \texttt{sunpy} package \citep{sunpy_community2020}. The algorithm follows the methodology introduced by \cite{TSC1985}, which we briefly outline here. The methodology allows us to estimate the temperature, $T$, and volume emission measure $EM=N_eN_HV$, where $N_e$ is the electron density, $N_H$ the proton density, and $V$ the homogeneous and isothermal source volume, from the reported X-ray fluxes $B_4$ ($0.5-4$\AA\,channel) and $B_8$ ($1-8$\AA\,channel). The ratio of the two channels, $R=\frac{B_4}{B_8}$, can be inverted to determine temperature because it is a monotonic function in the $1-100$ MK range. Once the temperature is known, the emission measure is derived from the longwave $1-8$\AA\,channel using a temperature-dependent scaling factor. The scaling between $R$ and $T$, as well as $B_8(T)$ and $EM$, is found by calculating GOES X-ray Spectrometer (XRS) responses in the two channels. To determine the temperature response of the reported GOES XRS fluxes, \cite{TSC1985} considered the measured X-ray flux values $B_i$, which depend on temperature and volume emission measure as
\begin{equation}
B_{i} = EM\int_{0}^{\infty}G_{i}(\lambda)f(T, \lambda)d\lambda/\overline{{G_{i}}},    
\end{equation}

\noindent where \(G_{i}(\lambda)\) is a wavelength-dependent transfer function, \(EM\) is volume emission measure,  $f(T, \lambda)$ is the isothermal emission spectra and \(\overline{{G_{i}}}\) is the wavelength-averaged transfer function. The transfer functions for each of the GOES detectors can be found through the data available at the
National Geophysical Data Center and/or the Solar Data Analysis Center, derived from the comprehensive calculations performed by \cite{Garcia1994}.

The derived temperature and emission measures are sensitive to whether coronal or photospheric abundances are assumed, even at higher temperatures, where continuum emission dominates over characteristic emission lines in the Sun's X-ray spectrum. This sensitivity occurs because a coronal abundance allows free-bound rather than free-free emission alone in the continuum \citep{White2005}. Nevertheless, this sensitivity does not alter the relationship between distributions of flare temperatures, as the majority of the events observed in this study are relatively low-temperature flares ($\sim$ 15 MK), which yield similar derived temperatures regardless of the assumed abundance. Using the derived polynomial approximations from \cite{TSC1985}, the temperature is obtained from 

\begin{equation}
T(R)=A(0)+A(1)R+A(2)R^2+A(3)R^3    
\end{equation}

\noindent and once the temperature is known, the emission measure is derived from $EM=10^{55}B_8/b_8(T)$ cm$^{-3}$, where $B_8$ is the $1-8$\AA\, flux in the standard GOES unit of W\,m$^{-2}$ and

\begin{equation}
b_8(T)=B(0)+B(1)T+B(2)T^2+B(3)T^3  
\end{equation}

\noindent is the normalized response. \cite{White2005} (and the \texttt{sunpy} code) derives the polynomial approximations for all GOES satellites using CHIANTI \citep{Landi1999,Landi2002} spectral models, which then gives us the flare temperature and the total amount of hot plasma in the flare regions (i.e., the emission measure) of the flare events.

\section{Results}\label{sec:results}

\begin{figure}[H]
\centerline{\includegraphics[width=1\textwidth,clip=]{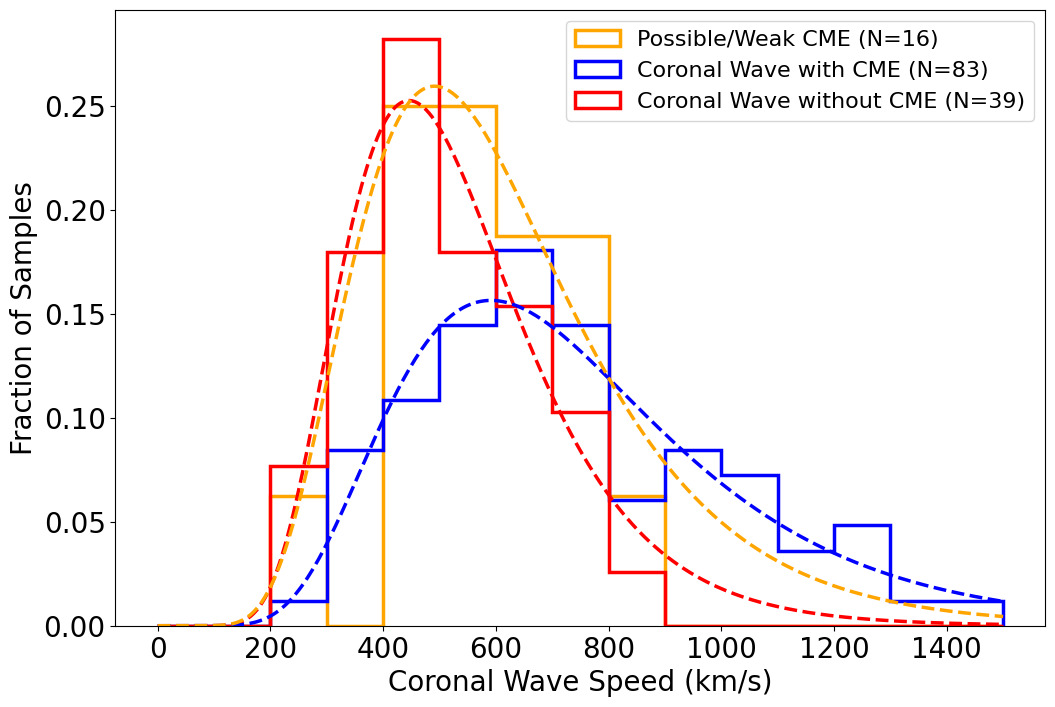}}
\caption{A histogram showing 171\,\AA\, EUV wave events (2010-2013) from the LCPF AIA movie catalog \citep{Nitta2014} were cross-referenced with the LASCO CME Catalog and observed for the presence of a CME using a combination of AIA and coronagraph data in 193\,\AA. The y-axis gives the fraction of EUV wave events in the catalog, and the $x$-axis highlights EUV wave speeds. The distribution in red includes EUV wave events that were not followed by a CME, the distribution in blue includes EUV wave events that were noticeably followed by a CME, and the orange distribution includes indeterminate events. The dashed lines indicate the corresponding fitted log-normal distributions from the observed mean and standard deviation.} \label{fig:LCPFspeed}
\end{figure}

First, we separate the distribution of LCPF propagation speeds based on the presence or absence of an accompanying CME (Figure \ref{fig:LCPFspeed}). Colors in Figure \ref{fig:LCPFspeed} highlight three different populations of coronal wave events: coronal waves with a CME (blue), coronal waves without a CME (red), and coronal waves that were difficult to identify whether or not they have a CME (orange). Since all coronal wave events correlate with flare events, we applied a log-normal fit to each histogram distribution \citep{Aoki2004,Verbeeck2019}. If the continuous random variable $X$ is log-normally distributed, then $Y=\ln(X)$ has a normal distribution. If $\mu$ and $\sigma$ are the mean and standard deviation of $Y$, then $X=\exp(\mu+\sigma Z)$, where $Z$ is a standard normal variable. The lognormal distribution also has mean $\exp(\mu+{\sigma^2}/{2})$, median $\exp(\mu)$, and variance $[\exp(\sigma^2)-1]\exp(2\mu+\sigma^2)$ \citep{crow1988lognormal}. The mean, median, and standard deviation, with uncertainties found using the standard error propagation formula, are given in Table \ref{tab:LCPFspeed} of the Appendix.

\begin{figure}
\centerline{\includegraphics[width=1\textwidth,clip=]{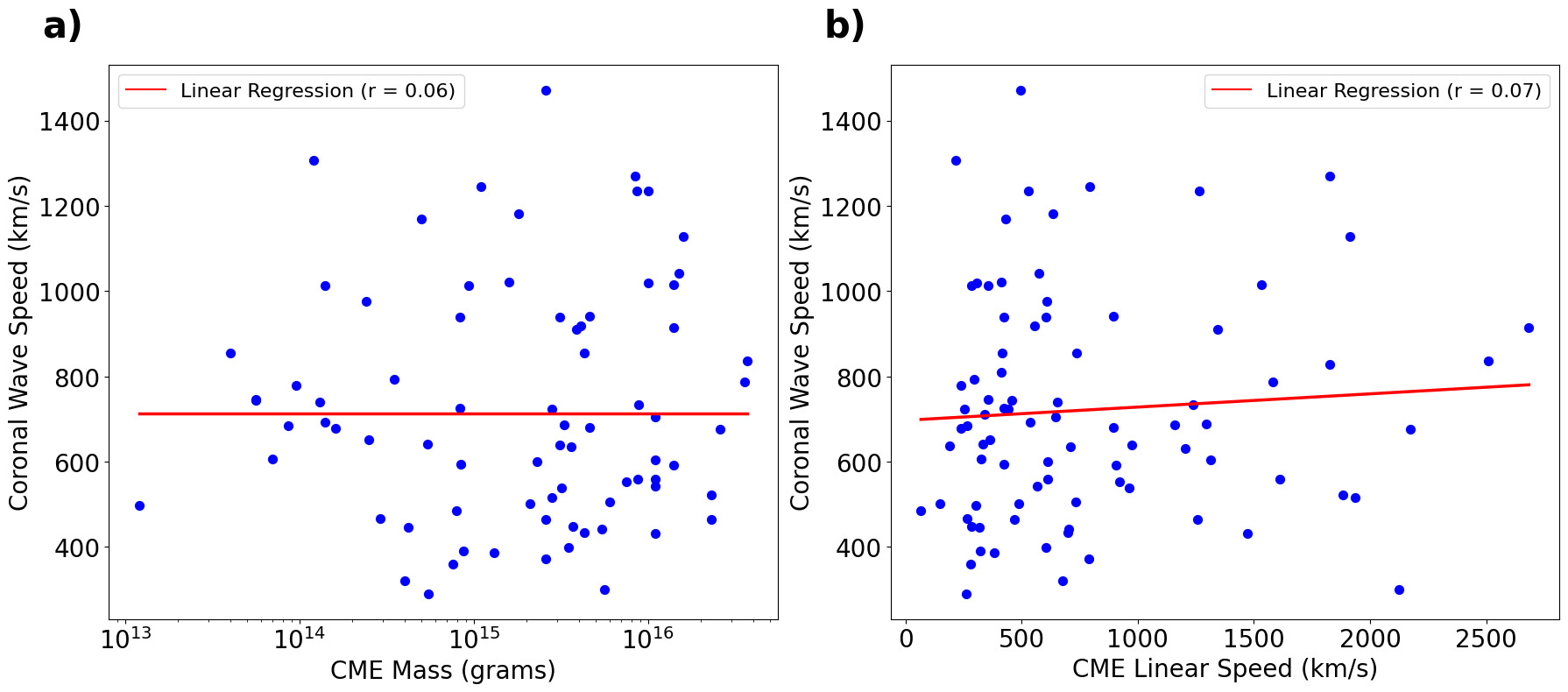}}
\caption{Scatter plot of CME mass versus coronal wave speeds (a) and CME linear speed versus coronal wave speeds (b). The data on the $x$-axis was obtained via the LASCO CME catalog, while the $y$-axis is the 2010-2013 coronal speed data from \citet{Nitta2013}}\label{fig:CMEmass+speed}
\end{figure}

We observed that the EUV waves had differing median propagation speeds depending on classification, where those that had an accompanying CME traveled with median speed $698.75 \pm 28.22$ km/s, those that did not have an accompanying CME traveled with median speed $504.15\pm 18.96$ km/s, and events where a visual inspection could not conclude the presence of a CME with median speed $572.32\pm 31.56$ km/s. From these distributions, we see that coronal waves without a CME are significantly slower and more concentrated between $300-700$ km/s, while the coronal waves with a CME are much more distributed across the axis, supported by its large standard deviation of $323.83\pm 44.71$ km/s. Then, we analyzed the correlation of the coronal wave events associated with a CME to that respective CME's linear mass and speed (Figure \ref{fig:CMEmass+speed}). We find only a weak correlation between coronal wave speed and CME mass, and an even weaker (though skewed) dependence on CME speed. The linear regression lines are very weak, and the Spearman correlation coefficients are 0.02 and 0.07, respectively.

\begin{figure}
\centerline{\includegraphics[width=1\textwidth,clip=]{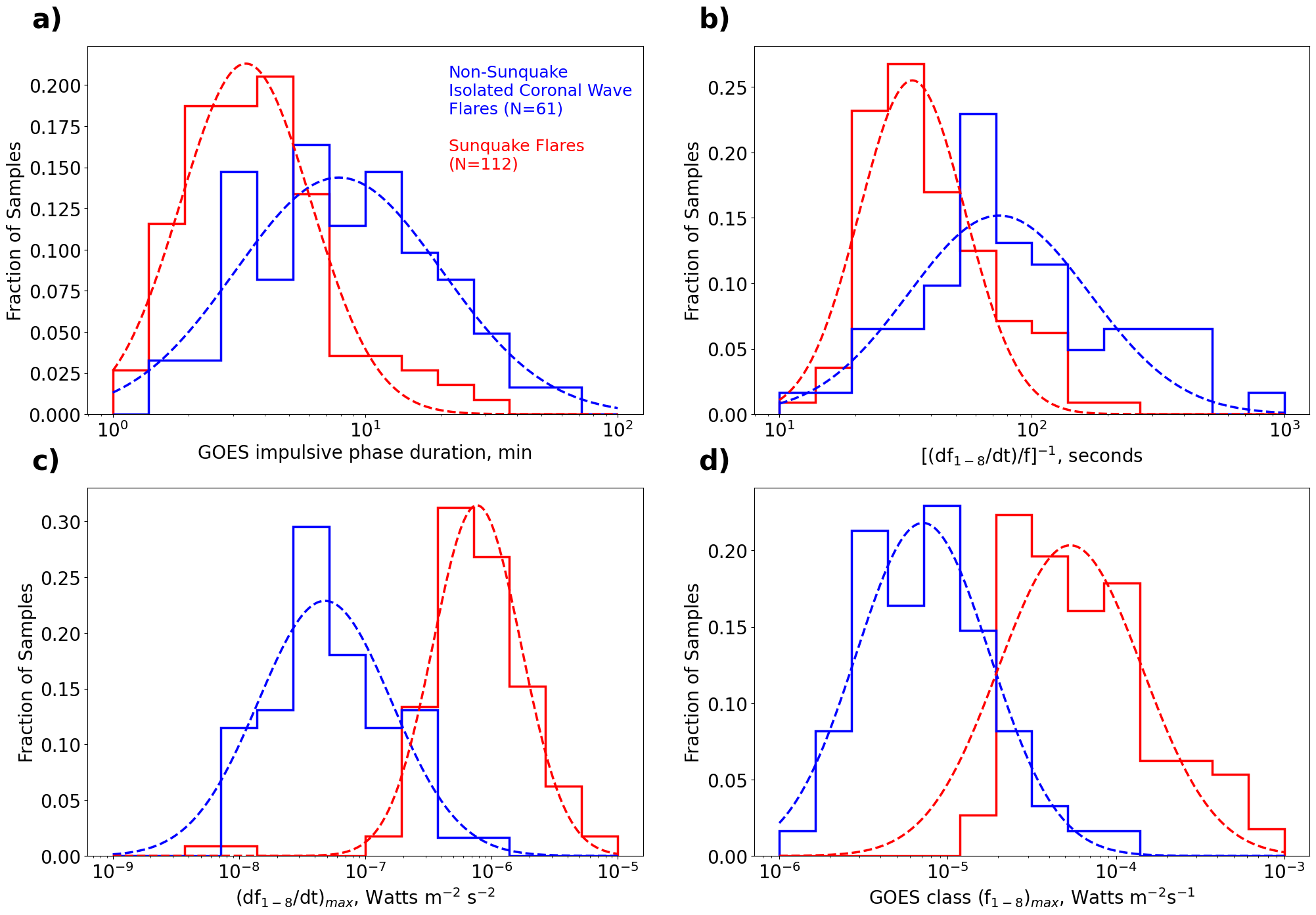}}
\caption{These four histograms compare certain flares of C1.0 or greater from 2010-2022. The red highlights flares that were associated with sunquakes as outlined in \citet{Sharykin2020}, while those in blue are coronal wave flares (without a CME or filament eruption) recorded in \cite{Nitta2013}. Panel a shows the impulsive phase duration, panel b shows the characteristic energy release time estimated as the maximum value of $f_{1-8\,\AA\,}/(df_{1-8\,\AA\,}/dt)$, panel c shows the maximum value of the SXR flux time derivative, and panel d shows the SXR flux. The symbol $f_{1-8\,\AA\,}$ refers to solar X-ray fluxes in the 1--8-Angstrom spectrum measured by GOES. }\label{fig:sunquakehistogram}
\end{figure}

Flare X-ray properties from the GOES data provide the most important parameters of the flare energetics, meaning the statistical comparison of the X-ray and helioseismic energetics
is an important step to understanding the physical processes associated with the helioseismic response. We first compare the population of isolated coronal wave flares to the population of sunquakes (Figure \ref{fig:sunquakehistogram}), as a potential co-excitation of coronal waves and sunquakes was the original motivation of this study. Colors in Figure \ref{fig:sunquakehistogram} distinguish two populations of flare events: flares with coronal waves lacking both an accompanying CME and a sunquake (blue), and flares with an associated sunquake regardless of CME occurrence (red). The sunquake distribution is qualitatively similar to the one derived in \cite{Sharykin2020} for the sunquake events of Solar Cycle 24 --- overall, flares associated with sunquakes are highly impulsive events. Their impulsive and characteristic energy release times exhibit strongly left-skewed distributions, indicating shorter time durations of their impulsive phases. Moreover, these flares range from GOES classes M to X, and generally display higher maximum values of the SXR flux time derivative, indicating powerful, energetic flares that rapidly release energy.

However, there are some slight differences from the distribution in \citet{Sharykin2020}. This may be for a few reasons, most notably choices of binning, limits on the axes, and that, for purposes of this paper, we considered all 114 cataloged events as sunquakes (regardless of whether they were considered a "sunquake candidate" due to their weak acoustic signals). The mean, median and standard deviation, with uncertainties found using the standard error propagation formula, are found in Table \ref{tab:impulsiveduration} for the impulsive phase, Table \ref{tab:energyreleaserate} for the characteristic energy release rate, Table \ref{tab:maxtimederivGOES} for the maximum value of the SXR flux time derivative, and Table \ref{tab:maxGOESflux} for the SXR flux/GOES class. Figure \ref{fig:sunquakehistogram}d revealed that isolated (not associated with a CME) coronal waves have a lower maximum SXR flux that varies between the high C and low M flare classes, spread out less than the sunquake event distribution. Moreover, we see in Figure \ref{fig:sunquakehistogram}a that flares with isolated coronal waves are much less impulsive, and can span from a few minutes to almost one hundred minutes. This is further supported by the characteristic energy-release time in Figure \ref{fig:sunquakehistogram}b, which shows a smaller, less pronounced peak that has a much wider variation than the sunquake population. This difference is more pronounced in terms of the maximum values of the SXR flux time derivative in Figure \ref{fig:sunquakehistogram}c, which shows that the isolated coronal wave flares have an order of magnitude lower max SXR flux time derivative than that of sunquakes.

\begin{figure}
\centerline{\includegraphics[width=1\textwidth,clip=]{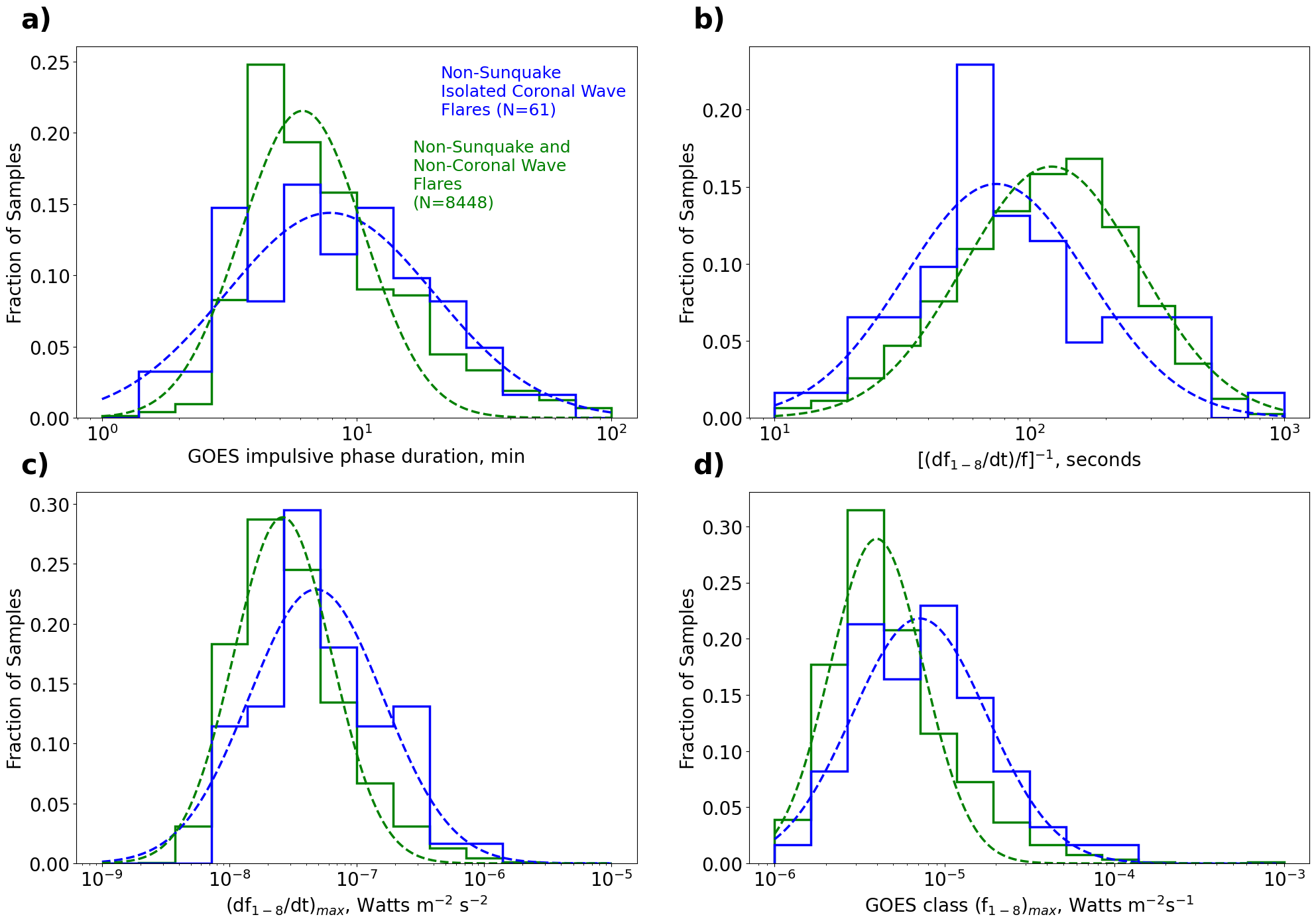}}
\caption{The distribution in blue corresponds to the same subset of isolated coronal wave flares in Figure \ref{fig:sunquakehistogram}, and the distribution in green represents all other flares that were not related to either sunquakes or coronal waves.}\label{fig:allflareshistogram}
\end{figure}

Then we compared flares in the 2010--2022 time frame that were not associated with a sunquake or coronal wave with the population of isolated coronal wave flares previously described (Figure \ref{fig:allflareshistogram}). Colors in Figure \ref{fig:allflareshistogram} highlight two populations of flare events: flares with coronal waves that did not have an accompanying CME and were not associated with a sunquake (blue), and flares that were not associated with any type of coronal wave or sunquake (green). Note that Tables \ref{tab:impulsiveduration}, \ref{tab:energyreleaserate}, \ref{tab:maxtimederivGOES} and \ref{tab:maxGOESflux} also have the mean, median, and standard deviations of these distributions as well. Figure \ref{fig:allflareshistogram}a shows that isolated coronal wave events (without sunquakes) have marginally higher impulsive phase durations than events without any photospheric or coronal seismic events. However, Figure \ref{fig:allflareshistogram}b indicates flares with isolated coronal waves still have a shorter characteristic growth time. Figure \ref{fig:allflareshistogram}c presents a slightly higher peak in the maximum SXR flux time derivative of the isolated coronal wave events, indicative of greater energy release rates. Figure \ref{fig:allflareshistogram}d shows that these events also have a consistently higher GOES flare class, meaning more intense radiation (and thus higher energy) than a flare with no sunquake or coronal wave. 

\begin{figure}
     \centering        \includegraphics[width=1\linewidth]{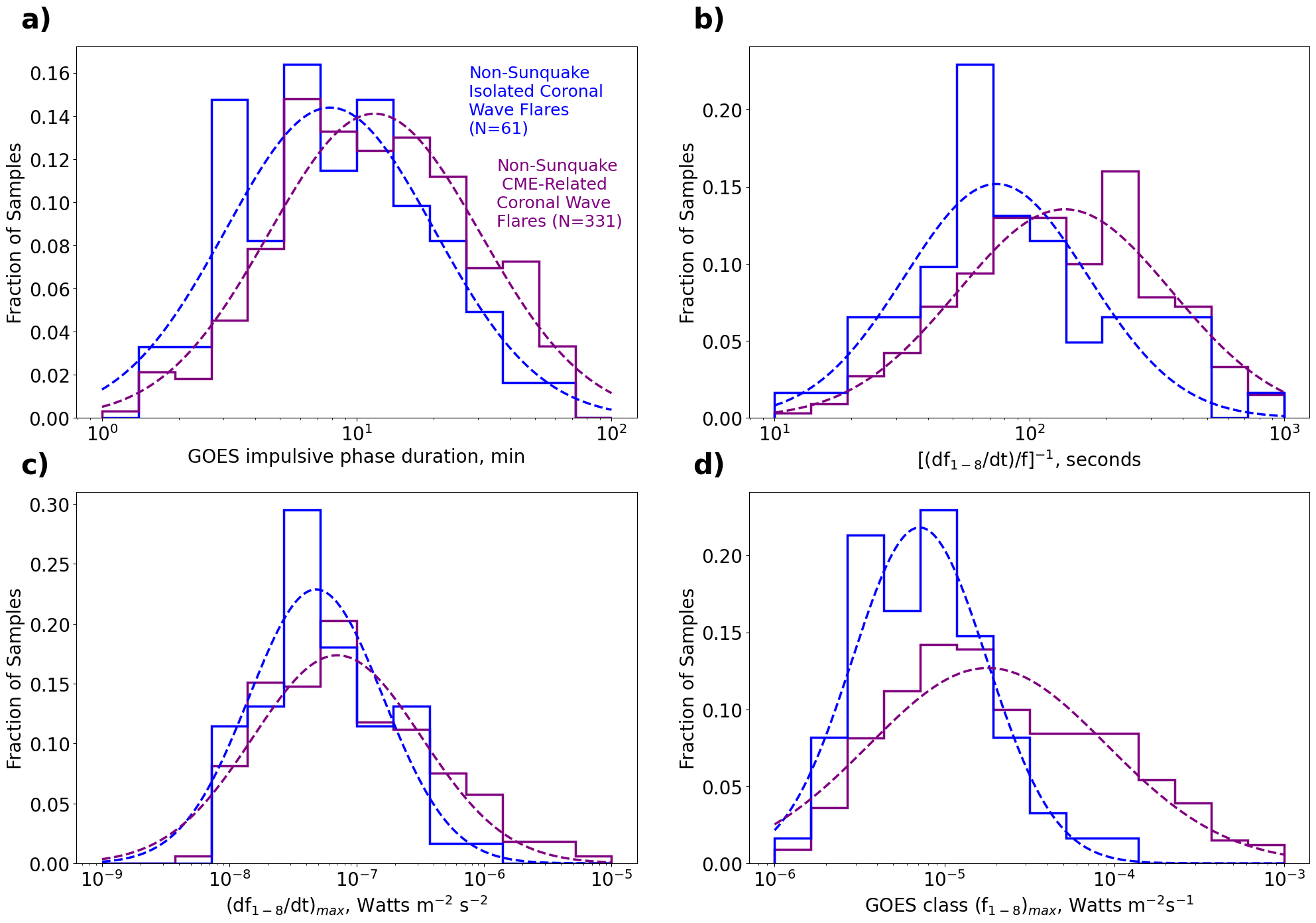}
    \caption{The distribution in blue corresponds to the same subset of isolated coronal wave flares in Figure \ref{fig:sunquakehistogram}, and the distribution in purple represents flares that do not have a sunquake, but are associated with coronal waves that have an accompanying CME.  }\label{fig:CMELCPFhistogram}
\end{figure}

Finally, we compare the same population of isolated coronal wave flares with the rest of the coronal wave population that was related to a CME (Figure \ref{fig:CMELCPFhistogram}). Colors in Figure \ref{fig:CMELCPFhistogram} highlight two populations of flare events: flares with coronal waves that did not have an accompanying CME and were not associated with a sunquake (blue), and flares with coronal waves that did accompany a CME and were not associated with a sunquake (purple). Again, Tables \ref{tab:impulsiveduration}, \ref{tab:energyreleaserate}, \ref{tab:maxtimederivGOES} and \ref{tab:maxGOESflux} also have the mean, median, and standard deviations of these distributions.

Overall, the CME-related distributions are much wider than the sunquake flares, non-coronal wave and non-sunquake flares, and isolated coronal wave flares, as given by the comparatively high standard deviation in Figures \ref{fig:CMELCPFhistogram}a--d. For example, in  Figure \ref{fig:CMELCPFhistogram}b, isolated (events without a CME) coronal-wave flare events have a standard deviation of $205.49\pm 126.35$ seconds compared to CME-related coronal wave flare events having a standard deviation of $697.70\pm 264.73$ seconds. Likewise in panel (d), isolated coronal wave flare events have a standard deviation of $2.85\times 10^{-5}\pm 1.28\times 10^{-5}$ Watts $m^{-2}s^{-1}$ while CME-related coronal wave events have a greater standard deviation of $3.12\times 10^{-3}\pm 3.12\times 10^{-3}$ Watts $m^{-2}s^{-1}$.  Figure \ref{fig:CMELCPFhistogram}a shows that isolated coronal wave flares have a shorter impulsive phase duration, meaning they are more impulsive and have faster energy release than flares that have CME-related coronal waves.  Figure \ref{fig:CMELCPFhistogram}b supports this conclusion, as the distribution of the CME-related coronal wave flares is shifted and has a much smaller peak. However, the two populations have similar distributions of maximum value of the SXR flux time-derivative---which characterizes the maximal energy release rate---shown in Figure \ref{fig:CMELCPFhistogram}c. Nevertheless, the isolated coronal wave flares' median is slightly lower than the CME-related coronal wave flares with a more pronounced peak in the distribution. Figure \ref{fig:CMELCPFhistogram}d presents CME-related coronal wave flares' GOES class as ranging from high C to X, with a median slightly higher than that of isolated coronal wave flares and most concentrated in the middle M class. Altogether, this reveals that while isolated coronal wave flares may have faster energy release, they release overall less energy than flares that are associated with a CME. 

Figure \ref{fig:emtemphistogram} presents distributions of the maximum volume emission measure and maximum temperature for each flare to illustrate further differences in magnetic energy release between the populations of CME-related and isolated coronal wave flares. The emission measure provides information about the density and volume of the hot plasma, and larger values (between \(10^{46}\) and \(10^{50}\) cm\({}^{-3}\)) are characteristic of stronger flares with more substantial heating.  When coupled with higher peak temperatures, these enhanced emission measures indicate a greater release of magnetic energy \citep{Feldman1996}. The temperature and emission measure of the flare events are computed using the GOES SXR flux as described in Section \ref{sec:tempEM}. Colors in Figure \ref{fig:emtemphistogram} highlight three populations of flare events (where sunquakes may be found in any of the three populations): flares with coronal waves that did not have an accompanying CME (red), flares with coronal waves that did accompany a CME (blue), and flare events without any coronal wave event (green). Moreover, the mean, median, and standard deviations of these distributions for the maximum emission measure histograms can also be found in Table \ref{tab:emphoto} using photospheric abundances and {Table \ref{tab:emcoronal} using coronal abundances. The same statistics are available for the flare temperatures in Table \ref{tab:tempphoto} using photospheric abundance and Table \ref{tab:tempcoronal} using coronal abundance.

Figures \ref{fig:emtemphistogram}a1 and \ref{fig:emtemphistogram}a2 show the maximum flare emission measure and temperature using the photospheric element abundance, while \ref{fig:emtemphistogram}b1 and \ref{fig:emtemphistogram}b2 use the coronal element abundance. The main difference between the two temperature estimates appears only to be a scaling factor, likely from FIP-related (first ionization potential effect) differences in the abundances; elements with FIPs of less than 10 eV have abundances three to four times greater in the corona than in the photosphere \citep{Doschek2018}. However, this difference is more pronounced in the emission measure, where cooler flares ($\sim15$ MK) have comparable values for the max flare temperature in both photospheric and coronal abundances; however, for hotter flares ($\sim35$ MK), coronal abundance leads to nearly $25$\% lower derived temperatures \citep{White2005}. 

\begin{figure}[H]
\centerline{\includegraphics[width=1\textwidth,clip=]{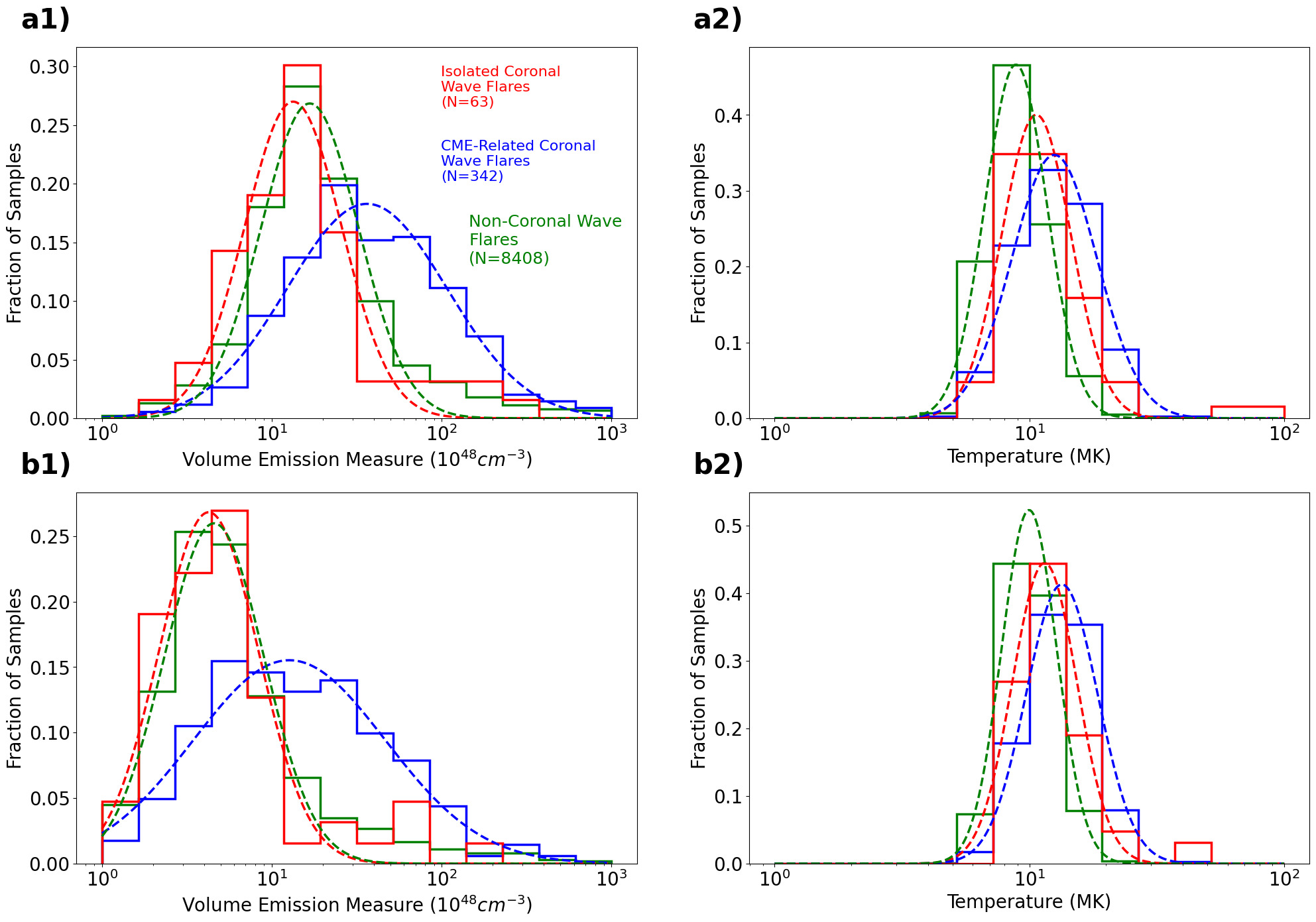}}
\caption{Histograms of emission measure and flare temperature, calculated with photospheric element abundances (a1 and a2, respectively) and with coronal abundances (b1 and b2, respectively), using code inspired by calculations from \cite{White2005}. The red distributions correspond to any coronal wave flare not accompanied by a CME, blue to flares with coronal waves that most likely have a CME, and those in green are the rest of the C1.0+ flares not considered in the other two categories.}\label{fig:emtemphistogram}
\end{figure}

Figures \ref{fig:emtemphistogram}a2 and \ref{fig:emtemphistogram}b2 show that CME-related coronal wave flares are the hottest, followed by isolated (without a CME) coronal wave flares and then by the other non-coronal wave flares. The non-coronal wave flare population's temperature distribution peaks (calculated for the photosphere element abundance) around 10 MK and has the most narrow distribution of the three populations. The widest temperature distribution is the CME-related flare population, which does not have a pronounced peak in calculations for either the photosphere element abundance or the coronal element abundance and is shifted to the right, signifying higher temperatures. The isolated coronal wave flare distribution has a clear peak calculated for the coronal element abundance that is higher than that of the non-coronal wave events, but is shifted to the left of the CME-related events. The uncertainties on the medians in Tables \ref{tab:tempphoto} and \ref{tab:tempcoronal} highlight that the CME-related coronal wave flare population has higher temperatures and is well-separated from the isolated coronal wave and non-coronal wave distributions. In other words, CME-related coronal wave flares have a more significant release of magnetic energy than isolated coronal wave flares because a higher temperature means more energy was released and converted into heat.

Panels (a1) and (b1) present a significant difference in the emission measure of coronal wave flares with and without a CME. Isolated coronal wave flares have the lowest emission measure and seem to fall in the same range as a general flare without any coronal wave event, while coronal wave events with a CME have significantly greater emission measure. It appears that this characteristic reveals a stark difference between the isolated coronal wave flares (median EM $6.69\times 10^{48}\pm 0.823\times 10^{48}$ cm$^{-3}$) and CME-related coronal wave flares (median EM $69.2\times 10^{48}\pm 15.9\times 10^{48}$ cm$^{-3}$); these numerical values are summarized in Table \ref{tab:emcoronal}. The broader distribution of CME-related coronal wave flares indicates a greater amount of hot, dense plasma, a key indicator of a more substantial and energetic flare event involving a significant amount of heated material that emits intense SXR radiation. This is supported by the nature of CMEs, which are ejections of large, dense volumes of hot plasma into the heliosphere. 

Moreover, we wanted to confirm the flare energy distribution for the three flare populations directly, analyzing whether the results agreed with findings from the distributions of the flare volume emission measure, temperature, and impulsivity (Figure \ref{fig:flareenergyhist}). We find that the energy released by isolated coronal wave flares (median $0.0247\times 10^{30}\pm$ $0.00884\times 10^{30}$ erg) is slightly greater than that of the non-coronal wave flares (median $0.0189\times 10^{30}\pm 0.00273\times 10^{30}$ erg), but both are one to two orders of magnitude lower than the energy released by CME-related coronal wave flares (median $1.41\times 10^{30}\pm 0.381\times 10^{30}$ erg); these numerical values are summarized in Table \ref{tab:flarenergy}. Similar to the trends in Figure \ref{fig:emtemphistogram}, the energy distribution of CME-related coronal wave flares is notably broader and spans a wide range of energies, while the isolated and non-coronal wave distributions exhibit more constrained, well-defined maxima.

\begin{figure}  \centerline{\includegraphics[width=1\textwidth,clip=]{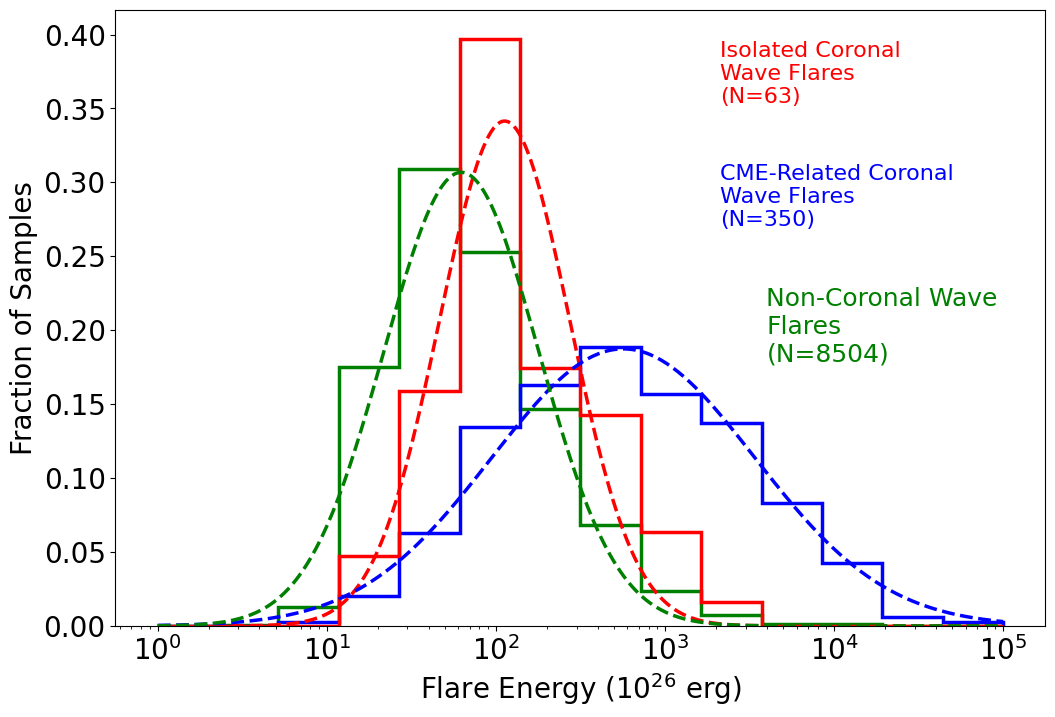}}
\caption{A histogram showing the flare energies for flares with GOES class C1.0 or greater from 2010-2022 (not including \textit{nan} values), calculated by integrating the area under the GOES light curve and then converting the units into erg.}\label{fig:flareenergyhist}
\end{figure}

We also investigated the correlation of the coronal wave speeds with both maximum flare emission measure and maximum flare temperature (using both coronal and photospheric abundances) to determine whether either of these factors influences the coronal wave speeds. Figures \ref{fig:emtempscatterplts}a1 and \ref{fig:emtempscatterplts}b1 show (with some outliers) that isolated coronal wave flares that occurred between 2010 and 2013 are generally slower and have a smaller emission measure than CME-related coronal wave flares. These CME-related coronal wave flares also have a weak linear correlation between wave speed and emission measure. Figures \ref{fig:emtempscatterplts}a2 and \ref{fig:emtempscatterplts}b2 demonstrate that the slower isolated coronal wave flares are also cooler and have an even weaker linear correlation between wave speed and temperature, as the temperature and speeds of CME-related coronal wave flares are much more sparsely populated. The dark, bright, or unclear fronts for each of these coronal wave events were observed by eye for each event recorded in \cite{Nitta2013} in 171\AA. The small sample of isolated coronal wave flares mostly have unclear or dark fronts, which seems to agree with both Figure \ref{fig:emtemphistogram} and Figures \ref{fig:emtempscatterplts}a2 and \ref{fig:emtempscatterplts}b2 in that these isolated wave events may be slower and cooler, thus having less energy. The emission measure versus wave speed seems to be slightly more correlated with wave speed than the temperature, with Figures \ref{fig:emtempscatterplts}a1 and \ref{fig:emtempscatterplts}b1 having Spearman correlation coefficients of 0.43 and 0.42 respectively, while \ref{fig:emtempscatterplts}a2 and \ref{fig:emtempscatterplts}b2 have correlation coefficients of 0.26 and 0.27 respectively.

\begin{figure}  \centerline{\includegraphics[width=1\textwidth,clip=]{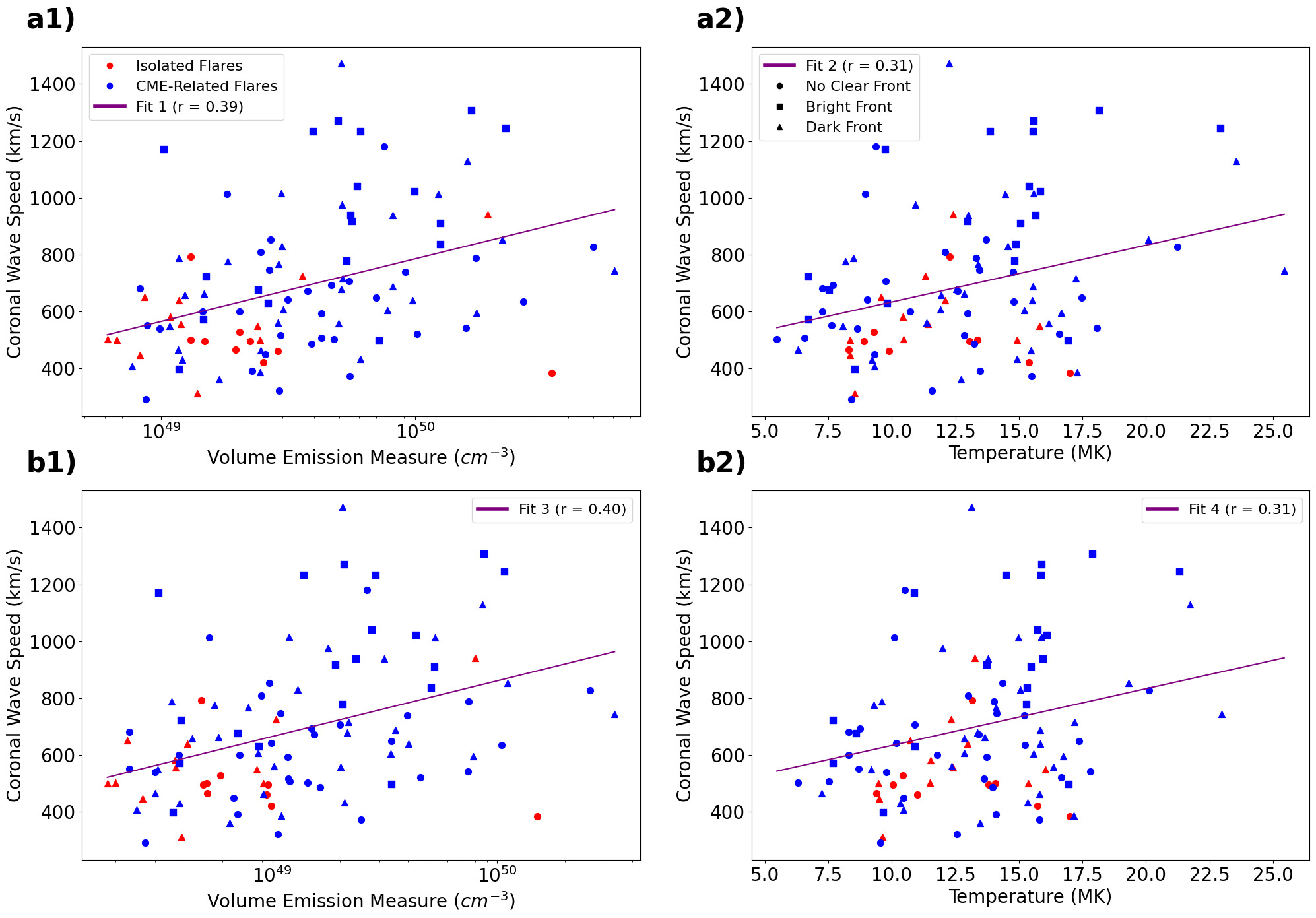}}
\caption{Scatter plots of emission measure versus coronal wave speeds calculated with the photospheric (a1) and coronal (b1) element abundances. Panels a2 and b2 show flare temperatures versus coronal wave speeds calculated with photospheric and coronal abundances, respectively. The data on the $x$-axis was calculated using the same code mentioned in Figure \ref{fig:emtemphistogram}, and the y-axis are the 2010-2013 events from \citet{Nitta2013}.} \label{fig:emtempscatterplts}
\end{figure}

\section{Discussion and Conclusion}\label{sec:conclusion}
We have performed a comprehensive study on solar flares from 2010-2022 that are correlated with coronal waves (or large-scale coronal propagating fronts) to understand how they may be affected by the presence of a coronal mass ejection. We analyze GOES SXR emission and its temporal evolution for coronal-wave-associated flares, comparing with events that have an accompanying CME as well as those without. The analysis further distinguishes non-sunquake isolated coronal wave flares from flares associated with sunquakes, and from flares exhibiting neither a sunquake nor a coronal wave. By cross-referencing catalogs of sunquakes, CMEs, and coronal waves, we statistically compare these flare populations. The main results can be summarized as follows:

\begin{enumerate}
    \item We find that coronal waves without an observed coronal mass ejection moved more slowly than coronal waves with a coronal mass ejection. Our fitting analysis has shown that coronal waves without CMEs are concentrated at lower propagation speeds with a clear maximum in the observed distribution, whereas coronal waves with CMEs span a much broader speed distribution that is skewed towards higher propagation speeds. The distribution of flare energies confirms flares that have CMEs are more energetic than those without; this suggests a correlation where higher energy release is associated with increased wave propagation speeds. Note that these wave speeds are originally from \citet{Nitta2013} and include only dates from 2010 to 2013.
    \item Our analysis also revealed that isolated flares with coronal waves release their magnetic energy faster (i.e. more impulsive) than CME-related coronal wave flares; however, CME-related coronal wave flares have more energy. This trend was initially indicated by higher maximum values of the soft X-ray time derivative and X-ray class, but  confirmed through temporal integration of the GOES SXR light curves for each recorded flare event. 
    \item Our finding that CME-related flares release more energy is further supported by their significantly higher maximum flare emission measure and slightly hotter maximum flare temperature compared to isolated coronal wave flares. This may influence the apparent, but weak, correlation between coronal wave speed and emission measure or temperature.
\end{enumerate}

In our initial investigation, we examined whether seismically active flares (those that produce sunquakes) can also generate coronal shockwaves from flare-accelerated particles. We also looked for similarities in the distributions of sunquake and isolated coronal wave flare events with the temporal characteristics of the flare soft X-ray emission. Although the results are not conclusive in answering the exact mechanism that generates either the flare's helioseismic response or coronal shock waves without a CME burst, it does support the conclusion that coronal waves without an accompanying CME are driven by a process distinct from the CME model scenarios, such as those described in \citet{Downs2011, Downs2012, Downs2021} and \citet{Cohen2009}.

The catalog from \citet{Nitta2013} contained only large-scale coronal propagating front speeds from 2010-2013, so further studies using speeds beyond 2013 with the time-distance method would allow us to draw better conclusions. Additionally, as mentioned by \cite{Sharykin2020}, a further detailed multi-spectral study considering characteristics of the
magnetic field structure and flare emission in other ranges of the electromagnetic spectrum would be promising, particularly in radio imaging. Type III radio bursts are associated with fast electron beams escaping the corona by traveling upward along magnetic field lines at near-relativistic speeds, exciting plasma emission that generates radio waves \citep{ReidRatcliffe2014}. The Expanded Owens Valley Solar Array \citep{Gary2018} observes in the microwave regime (1-18 GHz) with temporal resolution of about 1\,s at more than 100 frequency channels; moreover, they also provide solar spectral-imaging observations in high resolution, meaning it is quite easy to find the source region and the relationship between the burst and the magnetic field in that region \citep{Tan2023}. This can shed more light on what characteristics of flares, both in the pre-flare atmosphere and the flare dynamics, affect how energy is redistributed throughout the solar atmosphere. 

Overall, we find evidence that the generation of coronal waves without a CME is physically distinct from those generated with a CME. The substantially lower volume emission measure is a strong indicator distinguishing coronal wave events with or without associated CMEs. Isolated coronal wave events, or flares without a CME that are associated with a coronal wave event, are more impulsive (although weaker) than populations without coronal waves or CME-related coronal waves. While the distributions of isolated coronal wave flare properties differ from those of helioseismic events, the two populations share the important characteristic of rapid energy release. Finding further evidence of flare-accelerated particles' effect on the corona will allow for a unification of coronal seismology in the outer atmosphere with helioseismology and the propagation of acoustic waves into the Sun's interior. Moreover, it enables a more comprehensive characterization of solar flares and eventually more accurate prediction of space weather. 

\newpage

%%%%%%%%%%%%%%%%%%%%%%%%%%%%%%%%%%%%%%%%%%%%%%%%%%%%%%%%%%%%%%%%%%%%%%%%%%%
%% Appendix
%
\appendix
The following tables show the calculated median, mean and standard deviation
(with uncertainties found by error propagation) for the log-normal curve fits
of the plots in each figure. Note that the tables are separated by the certain
characteristic of the flare/coronal wave they highlight (i.e., coronal wave
speeds, impulsive phase duration, emission measure). Moreover, coronal and
photospheric abundances are divided into separate tables.

%% --- CME Wave Speed Statistics ---
\begin{table}[H]
\caption{Statistics of coronal wave speeds, in km\,s$^{-1}$, for different CME associations.}\label{tab:LCPFspeed}
\begin{tabular}{c c c c}
\hline
Category &
Median &
Mean &
Std. Dev. \\
\hline
Coronal wave without CME &
$504.15 \pm 18.96$ &
$536.16 \pm 23.81$ &
$194.08 \pm 27.60$ \\
Coronal wave with CME &
$698.75 \pm 28.22$ &
$759.60 \pm 37.17$ &
$323.83 \pm 44.71$ \\
Possible/weak CME &
$572.32 \pm 31.56$ &
$617.87 \pm 40.25$ &
$251.37 \pm 47.39$ \\
\hline
\end{tabular}
\end{table}

%Flare stats
% Impulsive phase duration Flare Statistics
\begin{table}[H]
\caption{Statistics of the impulsive phase duration in minutes for different flare populations.}\label{tab:impulsiveduration}
\begin{tabular}{c c c c}
\hline
Category &
Median &
Mean &
Std. Dev. \\
\hline
Isolated coronal wave &
$19.07 \pm 5.05$ &
$29.72 \pm 11.22$ &
$35.52 \pm 20.54$ \\
Sunquake &
$4.79 \pm 0.33$ &
$5.72 \pm 0.53$ &
$3.72 \pm 0.67$ \\
No sunquake or coronal wave &
$8.33 \pm 0.86$ &
$9.71 \pm 1.30$ &
$5.82 \pm 1.57$ \\
CME-related coronal wave &
$29.29 \pm 4.57$ &
$46.24 \pm 10.15$ &
$56.51 \pm 18.65$ \\
\hline
\end{tabular}
\end{table}

% Characteristic growth time  Flare Statistics
\begin{table}[H]
\caption{Statistics of the characteristic energy release time, $[df_{1-8}/dt/f]^{-1}$ in seconds for different flare populations.}\label{tab:energyreleaserate}
\begin{tabular}{c c c c}
\hline
Category &
Median  &
Mean &
Std. Dev. \\
\hline
Isolated coronal wave &
$147.21 \pm 40.81$ &
$207.28 \pm 79.74$ &
$205.49 \pm 126.35$ \\
Sunquake &
$42.47 \pm 3.97$ &
$47.71 \pm 5.66$ &
$24.43 \pm 6.44$ \\
No sunquake or coronal wave &
$229.37 \pm 12.81$ &
$314.11 \pm 24.15$ &
$293.88 \pm 36.77$ \\
CME-related coronal wave &
$351.09 \pm 63.09$ &
$560.56 \pm 141.99$ &
$697.70 \pm 264.73$ \\
\hline
\end{tabular}
\end{table}

% Max time derivative of GOES flux Flare Statistics
\begin{table}[H]
\caption{Statistics of the maximum time derivative of the GOES SXR flux, 
$(df_{1-8}/dt)_{\mathrm{max}}$ in $10^{-6}Wm^{-2}{s}^{-2}$ for different flare populations.}\label{tab:maxtimederivGOES}
\begin{tabular}{c c c c}
\hline
Category &
Median  &
Mean  &
Std. Dev. \\
\hline
Isolated coronal wave &
$0.209 \pm 0.124$ &
$0.436 \pm 0.383$ &
$0.798 \pm 1.000$ \\
Sunquake &
$1.450 \pm 0.138$ &
$2.000 \pm 0.261$ &
$1.890 \pm 0.401$ \\
No sunquake or coronal wave &
$0.0571 \pm 0.00537$ &
$0.0842 \pm 0.0110$ &
$0.0914 \pm 0.0186$ \\
CME-related coronal wave &
$0.717 \pm 0.350$ &
$2.290 \pm 1.650$ &
$6.950 \pm 6.790$ \\
\hline
\end{tabular}
\end{table}

% Max GOES SXR flux Flare Statistics
\begin{table}[H]
\caption{Statistics of the max GOES SXR flux, 
$(f_{1-8})_{\mathrm{max}}$ in $10^{-5}Wm^{-2}{s}^{-1}$ for different flare populations.}\label{tab:maxGOESflux}
\begin{tabular}{c c c c}
\hline
Category &
Median  &
Mean  &
Std. Dev. \\
\hline
Isolated coronal wave &
$1.64 \pm 0.341$ &
$2.49 \pm 0.729$ &
$2.85 \pm 1.28$ \\
Sunquake &
$14 \pm 5.87$ &
$22.8 \pm 13.9$ &
$29 \pm 27$ \\
No sunquake or coronal wave &
$0.595 \pm 0.0491$ &
$0.726 \pm 0.0801$ &
$0.509 \pm 0.104$ \\
CME-related coronal wave &
$24.2 \pm 12.1$ &
$88.4 \pm 65.3$ &
$312 \pm 312$ \\
\hline
\end{tabular}
\end{table}

% Volume emission measure of photospheric abundance Flare Statistics
\begin{table}[H]
\caption{Statistics of the max volume emission measure (photospheric abundance) in $10^{48}cm^{-3}$ for different flare populations.}\label{tab:emphoto}
\begin{tabular}{c c c c}
\hline
Category &
Median  &
Mean  &
Std. Dev. \\
\hline
Isolated coronal wave &
$20.5 \pm 2.36$ &
$25.5 \pm 3.91$ &
$18.7 \pm 5.16$ \\
CME-related coronal wave &
$117 \pm 17.2$ &
$211 \pm 44.1$ &
$315 \pm 95.2$ \\
Non-coronal wave &
$25.9 \pm 1.44$ &
$32.1 \pm 2.40$ &
$23.7 \pm 3.18$ \\
\hline
\end{tabular}
\end{table}

% Volume emission measure of coronal abundance Flare Statistics
\begin{table}[H]
\caption{Statistics of the max volume emission measure (coronal abundance) in $10^{48}cm^{-3}$ for different flare populations.}\label{tab:emcoronal}
\begin{tabular}{c c c c}
\hline
Category &
Median  &
Mean  &
Std. Dev. \\
\hline
Isolated coronal wave &
$6.69 \pm 0.823$ &
$8.38 \pm 1.40$ &
$6.35 \pm 1.89$ \\
CME-related coronal wave &
$69.2 \pm 15.9$ &
$162 \pm 54.1$ &
$342 \pm 159$ \\
Non-coronal wave &
$7.23 \pm 0.44$ &
$9.09 \pm 0.747$ &
$6.93 \pm 1.01$ \\
\hline
\end{tabular}
\end{table}

% Max Temperature of photospheric abundance Flare Statistics
\begin{table}[H]
\caption{Statistics of the max flare temperature (photospheric abundance) in megakelvin ($MK$) for different flare populations.}\label{tab:tempphoto}
\begin{tabular}{c c c c}
\hline
Category &
Median  &
Mean  &
Std. Dev. \\
\hline
Isolated coronal wave &
$11.73 \pm 0.41$ &
$12.32 \pm 0.51$ &
$3.94 \pm 0.53$ \\
CME-related coronal wave &
$14.54 \pm 0.30$ &
$15.67 \pm 0.39$ &
$6.27 \pm 0.42$ \\
Non-coronal wave &
$9.58 \pm 0.08$ &
$9.96 \pm 0.09$ &
$2.84 \pm 0.09$ \\
\hline
\end{tabular}
\end{table}

% Max Temperature of coronal abundance Flare Statistics
\begin{table}[H]
\caption{Statistics of the max flare temperature (coronal abundance) in $MK$ for different flare populations.}\label{tab:tempcoronal}
\begin{tabular}{c c c c}
\hline
Category &
Median  &
Mean  &
Std. Dev. \\
\hline
Isolated coronal wave &
$12.49 \pm 0.30$ &
$13.03 \pm 0.37$ &
$3.87 \pm 0.40$ \\
CME-related coronal wave &
$14.90 \pm 0.36$ &
$15.70 \pm 0.44$ &
$5.22 \pm 0.46$ \\
Non-coronal wave &
$10.62 \pm 0.06$ &
$10.96 \pm 0.07$ &
$2.78 \pm 0.09$ \\
\hline
\end{tabular}
\end{table}

% Max flare energy Flare Statistics
\begin{table}[H]
\caption{Statistics of the max flare energy in $10^{30}$ erg for different flare populations.}\label{tab:flarenergy}
\begin{tabular}{c c c c}
\hline
Category &
Median  &
Mean  &
Std. Dev. \\
\hline
Isolated coronal wave &
$0.0247 \pm 0.00884$ &
$0.0365 \pm 0.0183$ &
$0.0398 \pm 0.0309$ \\
CME-related coronal wave &
$1.41 \pm 0.3810$ &
$7.06 \pm 2.80$ &
$34.6\pm 18.3$ \\
Non-coronal wave &
$0.0189 \pm 0.00273$ &
$0.0329 \pm 0.00675$ &
$0.0468 \pm 0.0140$ \\
\hline
\end{tabular}
\end{table}

\clearpage

%%%%%%%%%%%%%%%%%%%%%%%%%%%%%%%%%%%%%%%%%%%%%%%%%%%%%%%%%%%%%%%%%%%%%%%%%%%
%% Acknowledgements
%
\begin{acks}
The observational data are courtesy of the SDO/AIA, GOES, and LASCO science teams. The CME catalog is generated and maintained at the CDAW Data Center by NASA and The Catholic University of America in cooperation with the Naval Research Laboratory. SOHO is a project of international cooperation between ESA and NASA. The LCPF catalog is compiled and maintained using SDO/AIA data by Nariaki V. Nitta at the Lockheed Martin Solar and Astrophysics Laboratory.
%The sunquake catalog was constructed by Ivan N. Sharykin at the IKI of the Russian Academy of Sciences and Moscow Institute of Technology as well as by Alexander G. Kosovichev at the New Jersey Institute Technology from SDO/HMI and GOES data. 
\end{acks}

%% Available additional data environments:
%% required: authorcontribution, fundinginformation, dataavailability
%% optional: materialsavailability, codeavailability
\begin{authorcontribution}
R.N. Bush wrote the main manuscript text and prepared figures 1-7 as well as tables 1-9. J.T. Stefan edited the main manuscript and assisted in the data collection and troubleshooting of figures 1, 3, 4, 5, and 6. A.G. Kosovichev edited the main manuscript and provided suggestions on error analysis for figures 1, 3, 4, 5, and 6. All authors reviewed the manuscript.
\end{authorcontribution}
\begin{fundinginformation}
The work was partially supported by the NSF grant 1916509, and NASA grants 80NSSC22M0162, 80NSSC23K0097, and 80NSSC25K7675.
\end{fundinginformation}
\begin{dataavailability}
The catalog of Large-scale Coronal Propagating Fronts (LCPFs) is available online at \url{https://aia.lmsal.com/AIA_Waves/index.html}, and the original paper reporting the 2010--2013 wave speeds is available at \url{https://doi.org/10.1088/0004-637X/776/1/58}. The catalog of sunquakes from Solar Cycle~24 is available at \\ \url{https://solarflare.njit.edu/sunquakes/sunquakes.html}, with the original paper available at \\ \url{https://doi.org/10.3847/1538-4357/ab88d1}. The SOHO LASCO CME Catalog is available at \url{https://cdaw.gsfc.nasa.gov/CME_list/}. The GOES data archive is available at \url{https://www.ngdc.noaa.gov/stp/satellite/goes/dataaccess.html}, with data analysis performed using SunPy documentation available at \url{https://docs.sunpy.org/en/stable/generated/gallery/time_series/goes_xrs_example.html}. Additionally, SunPy documentation describing the calculation of temperature and the corresponding volume emission measure from GOES/XRS data is available at \href{https://docs.sunpy.org/projects/sunkit-instruments/en/stable/api/sunkit_instruments.goes_xrs.calculate_temperature_em.html}{https://docs.sunpy.org/projects/sunkit-instruments/en/stable/api/sunkit\_instruments.goes\_x}\\
\href{https://docs.sunpy.org/projects/sunkit-instruments/en/stable/api/sunkit_instruments.goes_xrs.calculate_temperature_em.html}{rs.calculate\_temperature\_em.html}.
\end{dataavailability}
\begin{ethics}
\begin{conflict}
The authors declare no competing interests.
\end{conflict}
\end{ethics}

%%% %%%%%%%%%%%%%%%%%%%%%%%%%%%%%%%%%%%%%%%%%%%%%%%%%%%%%%%%%%%
%% Bibliography
%
% Using BibTeX
%
\bibliographystyle{spr-mp-sola} \bibliography{coronalwavebib.bib}

\end{document}